\documentclass[aps,pra,twocolumn, showpacs,amssymb,superscriptaddress]{revtex4-1}

\usepackage{graphicx, color}
\usepackage{amsmath,amssymb,amstext} 

\newcommand{\ket}[1]{\ensuremath{|#1\rangle}}
\newcommand{\bra}[1]{\ensuremath{\langle #1 |}}
\newcommand{\proj}[1]{\ket{#1}\!\bra{#1}}

\newcommand{\Let}{ \mathrel{\mathop : \! \! = } }

\newcommand{\norm}[1]{\lVert #1 \rVert}

\begin{document}
\title{Measuring geometric quantum discord using one bit of quantum information}

\author{G. Passante}
\affiliation{Institute for Quantum Computing and Dept.~of Physics and Astronomy, University of Waterloo, Waterloo, ON, N2L 3G1, Canada.}

\author{O. Moussa}
\affiliation{Institute for Quantum Computing and Dept.~of Physics and Astronomy, University of Waterloo, Waterloo, ON, N2L 3G1, Canada.}

\author{R. Laflamme}
\affiliation{Institute for Quantum Computing and Dept.~of Physics and Astronomy, University of Waterloo, Waterloo, ON, N2L 3G1, Canada.}
\affiliation{Perimeter Institute for Theoretical Physics, Waterloo, ON, N2J 2W9, Canada}

\date{\today}

\begin{abstract}
We describe an efficient DQC1-algorithm to quantify the amount of Geometric Quantum Discord present in the output state of a DQC1 computation. DQC1 is a model of computation that utilizes separable states to solve a problem with no known efficient classical algorithm and is known to contain quantum correlations as measured by the discord.  For the general case of a $(1+n)$-qubit DQC1-state we provide an analytical expression for the Geometric Quantum Discord and find that its typical (and maximum) value decreases exponentially with $n$.  This is in contrast to the standard Quantum Discord whose value for typical DQC1-states is known to be independent of $n$.  We experimentally demonstrate the proposed algorithm on a four-qubit liquid-state nuclear magnetic resonance quantum information processor. In the special case of a two-qubit DQC1 model, we also provide an expression for the Quantum Discord that only requires the outcome of the DQC1 algorithm. 
\end{abstract}


\maketitle


Since its inception in 2001, Quantum Discord (QD)~\cite{Henderson2001Classical-quant,Ollivier2002Quantum-Discord} and related measures have been used to quantify the amount of non-classical correlations in a physical system.  In addition to quantifying the most well known quantum correlations of entanglement, they also capture the quantum correlations that exist in separable states.  Discord measures are based on the premise that if a measurement on one part of a bipartite state disturbs the total state, there must be correlations stronger than what is found in the classical world.

Quantum discord has been studied extensively over the past few years, from operational definitions~\cite{Cavalcanti:2011ij,*Madhok:2011bs,*zurek2003quantum}, to witnesses~\cite{bylicka2010witnessing,*Laine:2010mi,*Ma:2011qa,*Maziero:2010kl,*rahimi2009single,*Smirne:2011pi,*Yu:2011fu,dakic2010necessary,Passante:2011kx} and analytical expressions for specific sets of states~\cite{Ali:2010ve,Ali:2010kx,Chen:2011dz,Girolami:2011ys,Lu:2011vn,Luo:2008uq}.  However, it is generally difficult to calculate as it requires both full state knowledge and an optimization over all projective measurements of a subsystem.  Even in the simple case of two qubits, a general closed form expression does not exist. The measure of Geometric Quantum Discord (GQD) was introduced~\cite{dakic2010necessary} as a simple geometric measure of the distance from a given state to the closest classical (zero discord) state. 

DQC1 is a model of mixed state quantum computation~\cite{Knill1998Power-of-One-Bi} that contains limited entanglement~\cite{Datta2005Entanglement-an} yet is thought to outperform classical methods. It does generate non-classical correlations as measured by the quantum discord~\cite{Datta2007Quantum-discord,Lanyon2008Experimental-qu,Passante:2011kx}, which, in hind sight, is not surprising since almost all quantum states have non-zero discord~\cite{Ferraro:2010fk}. Quantum discord has indeed been witnessed in the final states of DQC1 computations~\cite{Lanyon2008Experimental-qu}, even at very small values in highly mixed quantum states~\cite{Passante:2011kx}. To further explore the role of quantum correlations in the DQC1 model and beyond, their quantification becomes a necessary pursuit.
 
In this work, we derive an analytical expression for the geometric discord in the final state of a DQC1 computation of arbitrary dimension, in terms of quantities that can be efficiently estimated on a DQC1 computer. We then demonstrate its experimental evaluation in a four-qubit implementation of a DQC1 algorithm. Additionally, we derive a simple expression for the quantum discord in the special case of a two-qubit DQC1-state.

\section{Introduction}

Quantum discord is the most well-known measure of non-classical correlations and is thought to differentiate quantum and classical systems~\cite{Henderson2001Classical-quant,Ollivier2002Quantum-Discord}.  It is defined as the minimum difference between two classically equivalent formulations of the mutual information between subsystems $A$ and $B$, $I(A \mathop : B) \Let H(A) + H(B) - H(A,B)$ and $J(A \mathop : B) \Let H(B) - H(B|A)$, where $H(x)$ is the Shannon entropy when $x$ is described by a classical probability distribution, and is the von Neumann entropy when $x$ describes a quantum system.   In the quantum case the conditional entropy depends on the measurement basis, $H_{\{\Pi_k\}}(B|A) = \sum_k p_k H(\rho_{B|k})$, where $\{\Pi_k\}$ is a complete set of orthonormal projectors on $\cal{H}_A$ such that $\sum_k \Pi_k = \mathbb{I}$, $p_k$ is the probability of observing outcome $k$ on system $A$, and $\rho_{B|k} \Let \mbox{Tr}_A[(\Pi_k \otimes \mathbb{I}_B) \rho (\Pi_k \otimes \mathbb{I}_B) ]/p_k$ is the state of system $B$ conditional on the measurement of system $A$ returning measurement outcome $k$.  The Quantum Discord, $D(A\mathop : B)$, thus reduces to
\begin{equation}
D(A\mathop : B)= H(\rho_A) - H(\rho)  + \mbox{min}_{\{\Pi_k\}} \sum_k p_k H(\rho_{B|k}),
\label{discord}
\end{equation} 
where $\rho_A = \mbox{Tr}_B(\rho)$ is the reduced density matrix of system $A$.  In order to calculate this quantity, full state knowledge is required in addition to a minimization over all possible projective measurements on subsystem $A$.

While there is no analytical expression for the quantum discord of a general state, the discord for various sets of two-qubit states have been found~\cite{Luo:2008uq,Ali:2010kx,Chen:2011dz,Lu:2011vn}.  In addition, the quantum discord for all two-qubit states has been reduced to the problem of solving a set of transcendental equations~\cite{Girolami:2011ys}.  One of the results we report here is an expression for the QD in the final state of a two-qubit DQC1 algorithm that can be calculated with only the outcome of the DQC1 algorithm (trace of the unitary) and the initial polarization of the top register. 

Motivated by the difficultly in computing the QD, Dakic, Vedral and Brukner~\cite{dakic2010necessary} proposed the GQD, which is defined as
\begin{equation}
D_G^A(\rho) = \min_{\chi \in \Omega^A_0} \norm{\rho - \chi}^2\;,
\label{eqgqd_one}
\end{equation}
where $\Omega^A_0$ is the set of all zero discord states ($D(A\mathop :B)=0$.) These can be written as $\chi = \sum_j p_j \proj{j} \otimes \rho_j^B$.  The quantity $ \norm{ \rho - \chi }^2 = \mbox{Tr}(\rho - \chi)^2$ is the square of the Hilbert-Schmidt norm of Hermitian operators.  Note that the GQD, like the original quantum discord, is not symmetric in the subsystems, and $D_G^B(\rho)$ is defined as $\min_{\chi \in \Omega^B_0} \norm{ \rho - \chi }^2$ where $\Omega^B_0 = \sum_j p_j \rho_j^A \otimes \proj{j}$.

Shortly after the proposal of this measure of quantum correlations, Luo and Fu~\cite{Luo:2010ly} showed that Equation~\eqref{eqgqd_one} is equivalent to the minimization   
\begin{equation}
D_G^A(\rho) = \min_{\Pi^A} \norm{ \rho - \Pi^A(\rho) }^2,
\label{geod2}
\end{equation}
where $\Pi^A = \{ \Pi_k^A\}$ is a projective measurement on system $A$, and $\Pi^A(\rho) = \sum_k (\Pi^A_k \otimes I_B)\rho (\Pi^A_k \otimes I_B)$.   The minimization can be performed analytically for arbitrary $(2 \times 2)$-dimensional~\cite{dakic2010necessary} and $(2 \times d)$-dimensional~\cite{Vinjanampathy:2011ly} states.  The resulting expressions are in terms of quantities that are not efficiently experimentally accessible. However, a method for measuring a tight lower bound on the GQD has been recently proposed~\cite{Girolami:2011zr}, requiring only a constant number of measurements on up to four copies of the state.
In this report we provide an analytical expression for the GQD of $(2\times d)$-dimensional DQC1-states, and describe a DQC1-algorithm to efficiently estimate it.
\begin{figure}
\centering
\includegraphics[scale=0.4]{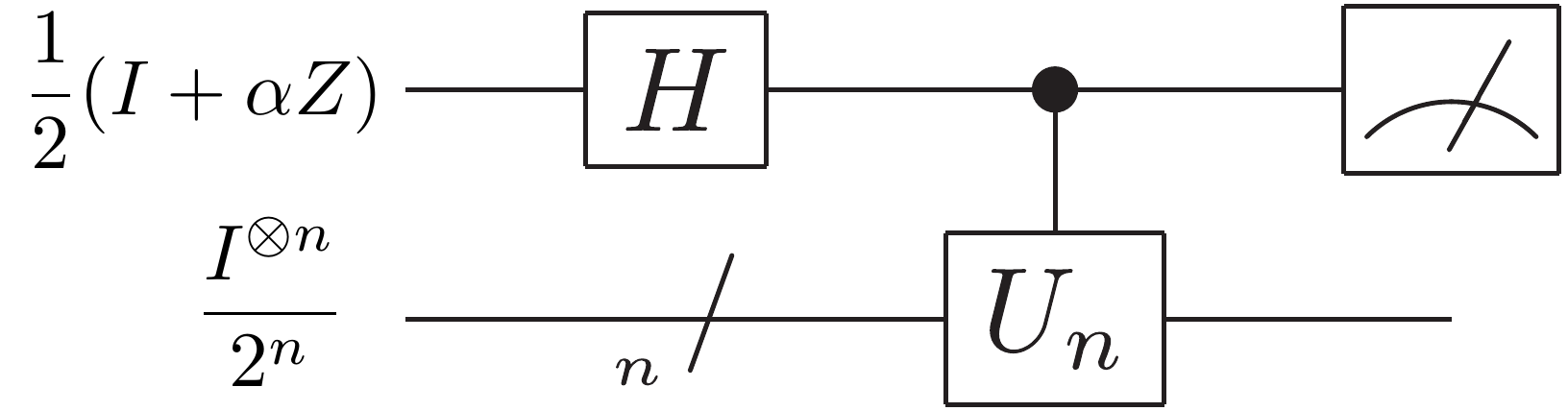}
\caption[DQC1 circuit for a mixed top register]{The circuit diagram above is for the DQC1 model of computation where the first qubit is in mixed state with polarization $\alpha$ along $\sigma_z$.  At the conclusion of the circuit, measurements of $\langle \sigma_x \rangle$ and $\langle \sigma_y \rangle$  result in the value of the real and imaginary components of the trace of the unitary, scaled by the value of the polarization of the first qubit.}
\label{DQC1fig}
\vspace{-.1in}
\end{figure}

The DQC1 model of computation is an example of a mixed state quantum information processor~\cite{Knill1998Power-of-One-Bi}.  As shown in Figure~\ref{DQC1fig}, this model has access to only one qubit with non-zero polarization, accompanied by a register of $n$ maximally mixed qubits. The computation consists of two gates: a Hadamard on the top, single qubit register, followed by an $n$-qubit unitary on the bottom register that is controlled by the state of the top qubit.  The final state of this simple algorithm is what we call the DQC1-state and can be written as
\[ \rho_{DQC1} = \frac{1}{2^{n+1}} \left( \begin{matrix} I^{\otimes n} & \alpha U_n^\dagger \\ \alpha U_n & I^{\otimes n} \end{matrix} \right). \]
Expectation value measurements of the Pauli matrices $\sigma_x$ and $\sigma_y$ yield the real and imaginary values of the trace of the unitary: $\frac{\alpha}{2^n}\mbox{Re}(\mbox{Tr}(U_n))$ and $\frac{\alpha}{2^n}\mbox{Im}(\mbox{Tr}(U_n))$.  Since calculating the trace of a unitary does not have a known efficient classical algorithm, a DQC1 computer is believed to be more powerful than its classical counterpart.

Despite its apparent improvement over classical methods, DQC1 is known to have zero bipartite entanglement between the top and bottom registers and can only have a small amount of entanglement across any other bipartite splitting~\cite{Datta2005Entanglement-an}.  It can, however, have quantum correlations characterized by the quantum discord~\cite{Datta2007Quantum-discord}.  This has been confirmed in two experiments to date: measurement of the discord in a two-qubit optics setup using full state tomography~\cite{Lanyon2008Experimental-qu}, and a four-qubit NMR implementation that witnesses discord with a small number of experiments~\cite{Passante:2011kx}.  

\section{Analytical expression for GQD}
In order to find an expression for the geometric discord in a DQC1-state, let us write Eqn.~\eqref{geod2} as
\begin{equation} 
\label{eqgqd_long}
D_G(\rho) = \min_{\Pi^A} \big(\mbox{Tr}(\rho^2) - 2\mbox{Tr}(\rho \Pi^A(\rho)) + \mbox{Tr}(\Pi^A(\rho)^2) \big).
\end{equation}
The first term, the purity of the state of the total system, is invariant under unitary transformations. Therefore it can be calculated for the initial state, and only depends on the initial polarization of the top register:
\begin{equation}
\mbox{Tr}(\rho_{DQC1}^2) = \frac{1+\alpha^2}{2^{n+1}}.
\end{equation}
In order to calculate the other two terms, consider parameterizing the measurement on the top qubit as $\Pi^A_\pm = \proj{\psi_\pm}$, where $\ket{\psi_+} = a\ket{0} + be^{i \phi}\ket{1}$ and $\ket{\psi_-} = b\ket{0} - ae^{i \phi}\ket{1}$; $a$ and $b = \sqrt{1-a^2}$ are real and $\in [0,1]$; and $\phi \in [-\frac{\pi}{2},\frac{\pi}{2}]$.  Thus, the state after measurement is
\begin{widetext}
\begin{equation}
\begin{split}
\Pi^A(\rho_{DQC1}) 
& = 
\left(\proj{\psi_+}\otimes I_n\right) \rho_{DQC1} \left(\proj{\psi_+}\otimes I_n\right) +  
\left(\proj{\psi_-}\otimes I_n\right) \rho_{DQC1} \left(\proj{\psi_-}\otimes I_n\right) \\
& = 
{ \frac{1}{2^{n+1}} \begin{pmatrix} I_n +\alpha ab(a^2-b^2)(e^{-i\phi}U+e^{i\phi}U^\dagger) & 2\alpha a^2b^2(e^{-2i\phi}U + U^\dagger) \\ 2\alpha a^2b^2(U + e^{2i\phi}U^\dagger) & I_n - \alpha ab(a^2-b^2)(e^{-i\phi}U+e^{i\phi}U^\dagger) \end{pmatrix} .}
\end{split}
\end{equation}
\end{widetext}

For the DQC1-state, the terms $\mbox{Tr}(\rho \Pi^A(\rho_{DQC1}))$ and $\mbox{Tr}( \Pi^A(\rho_{DQC1})^2)$ are equivalent, and evaluate to
\begin{equation*}
\frac{1}{2^{n+1}} + \frac{\alpha^2a^2b^2}{2^n}  + \frac{a^2b^2\alpha^2}{2^{2n}}  \sum_j \cos(2(\phi - \theta_j)),
\end{equation*}
where $\{\theta_j\}_{j=1... 2^n}$ are the eigenphases of $U$ (i.e. $U = \sum_{j=1}^{2^n} e^{i\theta_j} \proj{\theta_j}$), and we have used $\frac{1}{2} \mbox{Tr}(e^{2i\phi}U^{\dagger 2} + e^{-2i\phi}U^2) = \sum_j \cos(2(\phi - \theta_j))$. 
Thus, we are left with the task of minimizing
\begin{equation*}
\begin{split}
g(a,\phi;\{\theta_j\})
&=\norm {\rho_{DQC1} - \Pi^A(\rho_{DQC1}) }^2  \\
&= \frac{1+\alpha^2}{2^{2n+1}} - \frac{1 + 2\alpha^2a^2(1-a^2)}{2^{n+1}} \\
&\quad\quad- \frac{\alpha^2a^2(1-a^2)}{2^{2n}}  \sum_j \cos(2(\phi - \theta_j)),
\end{split}
\end{equation*}
over the measurement parameters $a \in [0,1]$ and $\phi \in [-\frac{\pi}{2},\frac{\pi}{2}]$. 
Examining first and second order derivatives of $g(a,\phi)$ with respect to $a$ and $\phi$, we find the optimal measurement parameters to be
\begin{eqnarray}
a_0 &=&  \frac{1}{\sqrt{2}}\textrm{, and} \\
\phi_0 
&=& \frac{1}{2} \arctan \left( \frac{\sum_j\sin(2 \theta_j)}{\sum_j\cos(2 \theta_j)} \right)\\
&=& \frac{1}{2} \arg(\textrm{Tr}(U^2)), \label{eqphi0}
\end{eqnarray}
which reduces the GQD to
\begin{equation}
\label{eqdgfinal}
D_G^A(\rho_{DQC1}) = \left(\frac{\alpha}{2}\right)^2 \frac{1}{2^{n}}\left[  1 -   \tau_2 \right],
\end{equation}
where $\tau_2 = |\mbox{Tr}(U^2)|/2^n$, and can be evaluated with a DQC1-algorithm using back-to-back applications of the control-$U$. The depth of the circuit that evaluates $\tau_2$ is at most double the one that evaluates $\mbox{Tr}(U)$, and therefore has the same efficiency of evaluation as the standard DQC1 algorithm. 
Worth noting here that $a_0=1/\sqrt 2$ indicates that the optimal measurement is always in the transverse (X-Y) plane of the Bloch sphere, independent of the implemented unitary. Also note that Eqn.~\eqref{eqphi0} allows for the ability to experimentally evaluate the optimal measurement parameter $\phi_0$ using the same data collected for the evaluation of $\tau_2$.

The GQD does not depend on the eigenvectors of the unitary, but rather the distribution of eigenphases. This gives rise to classes of unitaries that generate the same GQD in a DQC1 circuit. For instance, reproducing a result from Ref.~\cite{dakic2010necessary}, the set of unitaries that generate zero GQD (and hence, zero QD) will have $\tau_2=1$, implying $\mbox{Tr}(U^2) = e^{i \xi}2^n$ for some $\xi$. This is possible if and only if the eigenphases of $U$ are $\{\theta_j;\; 2\theta_j = \xi \pm 2\pi \;  \forall \; j\}$. That is to say  
$U = e^{i\xi/2}A$, where $A$ is a binary observable ($A^2 = I$). 

For a unitary drawn randomly according to the Haar measure, its eigenphases are randomly distributed over the unit circle~\cite{Diaconis:2003kl}. Thus, for large $n$, $\tau_2$ approaches zero and $D_G^A(\rho_{DQC1})$ approaches its maximum value of $(\frac{\alpha}{2})^2\ 2^{-n}$, i.e. the GQD of a DQC1-state decreases exponentially with the number of qubits.  This is in contrast to the average QD of a DQC1-state, which was shown in Ref.~\cite{Datta2007Quantum-discord} to be independent of the number of qubits for large $n$.

\section{Experimental measurement of the geometric discord}
We measure the quantum correlations, as quantified by the GQD, in a liquid-state nuclear magnetic resonance implementation of the DQC1 algorithm.  The qubits in NMR are ensembles of spin-1/2 nuclei. In this experiment, we use the four carbon-13 nuclei in the molecule trans-crotonic acid (molecular information and Hamiltonian parameters can be found in Ref.~\cite{passante2009experimental}.)
The experiment is implemented in a 16.7 T magnetic field Bruker Avance spectrometer,
where, for carbon-13 nuclei ($\gamma = 6.728284 \times 10^7 \mbox{ rad } \mbox{T}^{-1} s^{-1}$) at room temperature, the initial polarization of the thermal state is $\alpha = 1.4\times 10^{-5}$.
The spins are manipulated with radio frequency pulses numerically generated using the GRAPE algorithm~\cite{Khaneja2005Optimal-control,Ryan:2008fk}. The pulses are designed to have a fidelity of no less than 0.998 and are adjusted for non-linearities in the pulse generation and transmission by placing a pickup coil at the location of the sample and running a feedback loop to iteratively adjust the pulse shape for optimal transmission.    

\begin{figure}
\begin{center}
\includegraphics[scale=0.88]{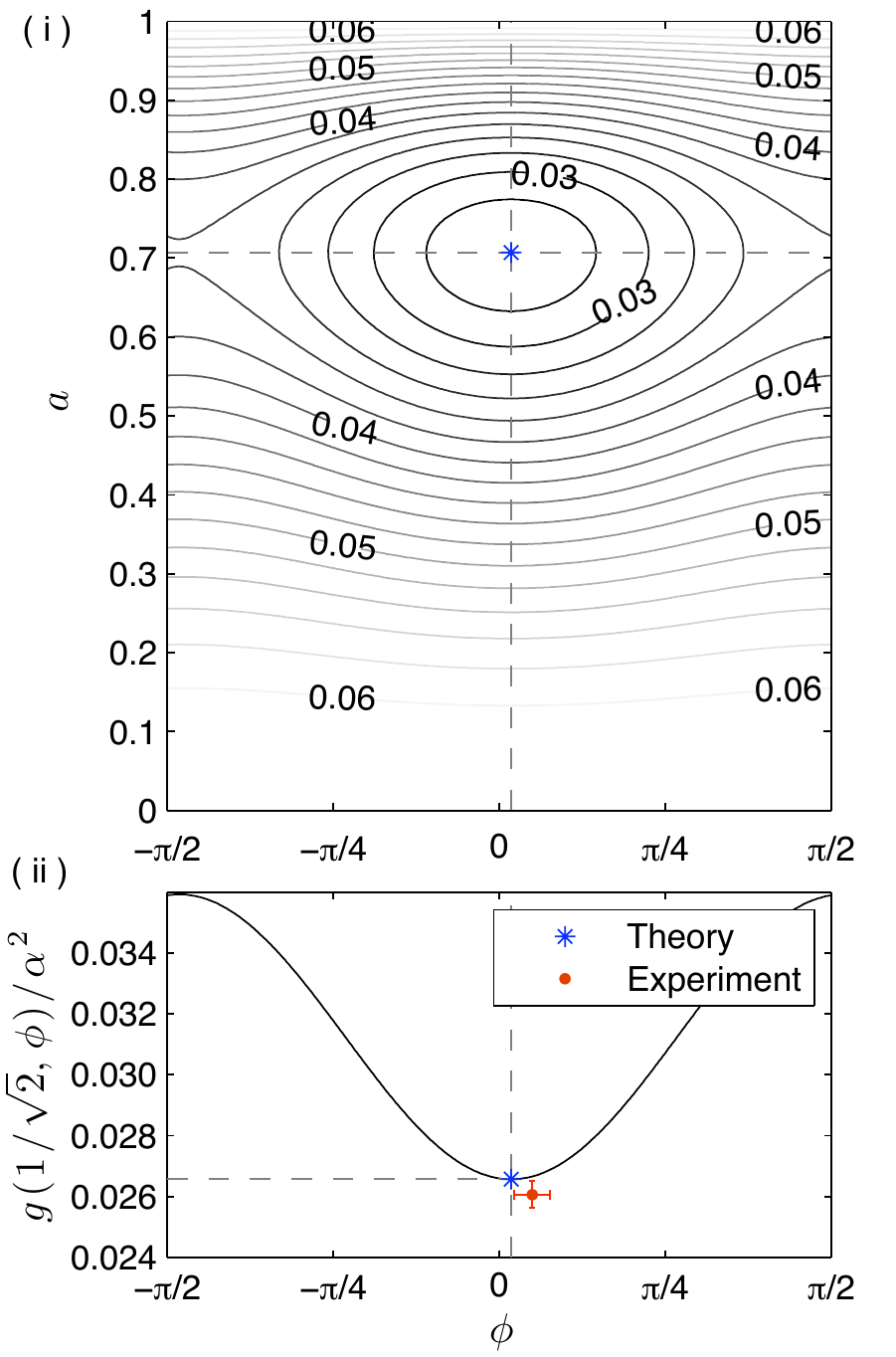}
\caption{(Color online) Shown are (i) the contour plot of the geometric distance, $g(a,\phi) =\norm {\rho_{DQC1} - \Pi^A(\rho_{DQC1}) }^2$, normalized by $\alpha^2$, as a function of the measurement parameters $a$ and $\phi$, for $\rho_{DQC1}$ the output of the four-qubit DQC1 algorithm described in the text, and (ii) the geometric distance at the optimal measurement axis $a= a_0 = 1/\sqrt{2}$. The dashed lines indicate the parameters that correspond to the optimal measurement. The experimental data point for the geometric discord of $\rho_{DQC1}$ is shown on plot (ii) with error bars propagated from experimental uncertainties and the spectral fit.}
\label{fig_results}
\end{center}
\vspace{-.1in}
\end{figure}

For the particular instance of the DQC1 circuit in this experiment, we choose to implement one of the unitary matrices used for approximating the Jones polynomial for braids with four strands~\cite{passante2009experimental}.  This is a problem that completely encapsulates the power of the DQC1 model and is of interest in many fields of physics and math. The unitary has the form $U = \mbox{diag}(c,c,d,1,c,d,1,1)$, where $c = -(e^{-i3\pi/5})^4$ and $d = (e^{-i3\pi/5})^8$. The DQC1-state, after the application of the control-$U$, is known to have nonzero discord as witnessed in a recent experiment~\cite{Passante:2011kx}. Using Eqn.~\eqref{eqdgfinal}, we calculate the expected value of the GQD to be $0.0266\, \alpha^2$. As shown in Figure~\ref{fig_results}, this is achieved for an optimal measurement with $a_0=1/\sqrt2$ and $\phi_0=0.116$~rad. Also shown in Figure~\ref{fig_results}(ii) is 
the experimental value of GQD in the DQC1-state, which is found to be $(0.0260\pm0.0004)\,\alpha^2$ by measuring the outcome of a DQC1 circuit with back-to-back applications of control-$U$. 
 For our experimental polarization ($\alpha = 1.4 \times 10^{-5}$), the GQD in the final state is $(5.10 \pm 0.08 )\times 10^{-12}$.

\section{The quantum discord of two-qubit DQC1-states}

In the case where the bottom register is a single qubit, we report an analytical expression for the QD of the DQC1-state, thereby contributing to the sets of two-qubit states for which the QD has been analytically solved. Commencing from Eqn.~\eqref{discord}, a minimization must be performed over the conditional entropy term, $\sum_k p_k H(\rho_{B|k})$, which, in the case of a two-qubit DQC1-state, reduces to
\begin{eqnarray*}
f(x) =  \sum_{j=1}^2  &-&\frac{1}{4} \log \left( \frac{(1/4+x_j)(1/4-x_j)}{p_+p_-} \right) \\* + \sum_{j=1}^2 &-& x_j \log \left( \frac{(1/4 + x_j)p_-}{(1/4 - x_j) p_+} \right),
\end{eqnarray*}
where $x_j = \tfrac{1}{2}\alpha a\sqrt{1-a^2}\cos(\phi - \theta_j)$, $p_\pm = 1/2 \pm \sum_j x_j$, and $a, b,$ and $\phi$ parameterize the measurement on the top qubit (see the GQD case above). The minimization with respect to the measurement parameters $a$ and $\phi$ results in an optimal measurement characterized by
\begin{displaymath}
a_0  = \frac{1}{\sqrt{2}} \;\; \mbox{and}  \;\;\; \phi_0 = \frac{\pi}{2} + \frac{\theta_1 + \theta_2}{2} .
\end{displaymath}
This allows us to write the analytical expression for the QD of a two-qubit DQC1-state as
\begin{equation}
\begin{split}
D(A \mathop: B) =  
&H_2 \left( \frac{1-\alpha\, \tau_1}{2}\right)  - H_2 \left(\frac{1-\alpha}{2}\right) \\
& - \frac{1}{2} \log \left( 1 -\alpha^2  (1 - \tau_1^2) \right) \\
&- \frac{\alpha}{2}  \sqrt{1 - \tau_1^2}\log \left( \frac{1 + \alpha  \sqrt{1 - \tau_1^2}}{1 - \alpha  \sqrt{1 - \tau_1^2}} \right),\\
\label{dqc1_D}
\end{split}
\end{equation}
where $H_2(\cdot)$ is the binary entropy, and $\tau_1 = |\mbox{Tr}(U)|/4$, which can be measured directly with a DQC1-algorithm.

Hence, the quantum discord of a two-qubit DQC1-state is zero for $|\mbox{Tr}(U)|^2=0$ or $4$, and maximum for $|\mbox{Tr}(U)|^2=2$. Recall, from Eqn.~\eqref{eqdgfinal}, that the GQD is zero for $ |\mbox{Tr}(U^2)| = 1/2^n$ and maximum for $ |\mbox{Tr}(U^2)| = 0$. 
Consider a general single-qubit unitary, expressed as $U=e^{i\xi}R_{\hat{r}}(\vartheta)$, where $R_{\hat{r}}(\vartheta):=\exp[-i \vartheta\, \hat{r}\cdot \vec{\sigma}/2]$ is a rotation by $\vartheta$ about the unit vector in three dimensions $\hat{r}=(r_x,r_y,r_z)$, $\vec{\sigma}=(\sigma_x,\sigma_y,\sigma_z)$ are the Pauli matrices, and $\xi$ a global phase. It follows from $(\hat{r}\cdot\vec{\sigma})^2=I$ that $\mbox{Tr}(U)=2e^{i\xi} \cos(\vartheta/2)$ and $\mbox{Tr}(U^2)=2e^{i2\xi} \cos(\vartheta)$. Thus, the QD and GQD are simultaneously maximum for $\vartheta= (2k+1)\pi/2$, and simultaneously zero for $\vartheta = k\pi$, independent of the rotation axis $\hat{r}$. Examples of unitaries that can be written as $\pi$ rotations include bit-flip (or NOT) and Hadamard.

\section{Conclusion}

In this article we provided analytical expressions for the quantum correlations present in a DQC1-state in terms of experimentally accessible quantities.  The geometric quantum discord can be computed for $(2\times d)$-dimensional DQC1-states by implementing a DQC1 algorithm with a second application of the controlled unitary.  The algorithm was experimentally demonstrated for a four-qubit liquid-state NMR implementation of the DQC1 model.  We also showed, that for the special case of a two-qubit system, the quantum discord can be calculated using outcome of the DQC1-circuit (the trace of the unitary). 

For a typical DQC1-state, where the unitary is chosen uniformly at random according to the Haar measure, we found that the geometric measure of discord 
scales with $2^{-n}$, in contrast to the quantum discord, which is known to be independent of $n$ (for large $n$). 
This suggests that the geometric quantum discord fails to completely quantify the correlations defined in the original entropic measure of quantum discord.  Finally, the work we have presented here further supports the suspicion that the apparent speedup exhibited by (the dynamics of) DQC1 is not necessarily captured by the geometric measure of quantum discord at the conclusion of the algorithm.


\begin{acknowledgments}
G.P.~would like to thank M.~Ditty for his technical expertise with the spectrometer.  This work was funded by NSERC, QuantumWorks, and CIFAR.  
\end{acknowledgments}


%

\end{document}